\documentclass{PoS}

\title{On the phase diagram of the Higgs \(SU(2)\) model}

\ShortTitle{On the phase diagram of the Higgs \(SU(2)\) model }

\author{\speaker{Claudio Bonati}\\
        Dipartimento di Fisica \& INFN, Pisa, Italy \\
        E-mail: \email{bonati@df.unipi.it}}

\author{Guido Cossu\\
        Dipartimento di Fisica \& INFN, Pisa, Italy \\
        E-mail: \email{g.cossu@sns.it}}

\author{Alessio D'Alessandro\\
        Dipartimento di Fisica \& INFN, Genova, Italy \\
        E-mail: \email{adales@ge.infn.it}}

\author{Massimo D'Elia\\
        Dipartimento di Fisica \& INFN, Genova, Italy \\
        E-mail: \email{delia@ge.infn.it}}

\author{Adriano Di Giacomo\\
        Dipartimento di Fisica \& INFN, Pisa, Italy \\
        E-mail: \email{digiacomo@df.unipi.it}}

\abstract{The Higgs \(SU(2)\) model with \(\lambda=\infty\) (fixed Higgs length) is usually believed to have two different phases at high 
gauge coupling \(\beta\), separated by a line of first order transitions but not distinuguished by any typical symmetry associated with a 
local order parameter, as first proved by Fradkin and Shenker. We show that in regions of the parameter space where it is usually supposed 
to be a first order phase transition only a smooth crossover is in fact present.}

\FullConference{The XXVI International Symposium on Lattice Field Theory \\
July 14 - 19, 2008\\
Williamsburg, Virginia, USA}

\begin{document}

\section{Introduction and motivation}

While in pure gauge theories there is a standard way to detect color confinement (the Wilson criterium \cite{Wilson}), when matter in the 
fundamental representation of the gauge group is present, large Wilson loops never obey the area-law, so that there is no obvious way
to clearly define the meaning of ``confined phase''. 

From the computational point of view, the simplest such model is the Higgs \(SU(2)\) model; to study 
only the general features of its phase diagram, this model can be simplified a bit more, fixing the length of the Higgs field. This is
possible because the scalar fourth-coupling \(\lambda\) is irrelevant in continuum limit and fixing the length of the Higgs field is 
equivalent to use \(\lambda=\infty\) (the irrelevance of \(\lambda\) was numerically checked in \cite{Montvay}). In this case the action
can be written in the form (see \emph{e.g.} \cite{Montvay})
\begin{equation}\label{azione}
S=\beta\sum_{x, \mu<\nu}\left\{1-\frac{1}{2}\mathrm{Re}\mathrm{Tr} P_{\mu\nu}(x)\right\}-
\frac{\kappa}{2}\sum_{x, \mu>0}\mathrm{Tr}[\phi^{\dag}(x)U_{\mu}(x+\hat{\mu})\phi(x+\hat{\mu})]
\end{equation}
where the first term is the standard Wilson action and the Higgs field \(\phi\) is written using an \(SU(2)\) matrix. This form of the 
action is particularly useful because with it standard heatbath (\cite{Creutz80}, \cite{KennedyPendleton}) and 
overrelaxation (\cite{Creutz87}) algorithms can be used to generate Monte Carlo configurations.

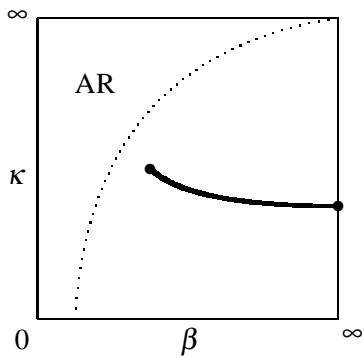
\begin{figure}[h]
\begin{center}
\setlength{\unitlength}{1cm}
\begin{picture}(4, 4)
 \put(0, 0){\line(0, 1){4}}
 \put(0, 0){\line(1, 0){4}}
 \put(0, 4){\line(1, 0){4}}
 \put(4, 0){\line(0, 1){4}}
 \put(4, 1.5){\circle*{.15}}
 \put(1.5, 2){\circle*{.15}}
 \put(1.9,-.4){\(\beta\)}
 \put(-.3,-.4){\(0\)}
 \put(4.05,-.3){\(\infty\)}
 \put(-.4,1.8){\(\kappa\)}
 \put(-.4,3.95){\(\infty\)}
 \put(0.5,3){AR}
 \qbezier[50](.5, 0)(.7, 3.5)(4, 4)
 \linethickness{0.5mm}
 \qbezier(1.5, 2)(2, 1.5)(4, 1.5)
\end{picture}
\end{center}
\caption{Phase diagram of the Higgs \(SU(2)\) model as predicted in \cite{FradkinShenker}.}
\label{diagramma}
\end{figure}

A theory with action \ref{azione} has the following important limiting cases
\begin{description}
\item[\(\kappa=0\)] : pure gauge theory, no transitions in \(\beta\)
\item[\(\beta=\infty\)] : \(O(4)\) non linear sigma model, it has a second order phase transition in \(\kappa\)
\end{description}
The general case was studied in \cite{FradkinShenker} using both perturbative and non-peturbative methods: using a pertubative expansion 
it was shown that the transition of the \(O(4)\) model is not lifted out by the introduction of the gauge field, but it becames a first 
order \emph{\`a la} Coleman-Weinberg. On the non-perturbative side, using the methods developed in \cite{OsterwalderSeiler}, the authors of
\cite{FradkinShenker} were able to prove the existence of a wide region of parameter space where \emph{every local observable is 
analytic}, the so called Fradkin Shenker (FS) theorem. Using these two inputs they suggested a phase diagram like that shown in Fig. 
\ref{diagramma}: the region where the analyticity is rigorously proven is indicated by AR and is limited by the dotted line, the thick 
line represents a line of first order transitions and the two dots are its second order end-points.
\newline
Since the FS theorem ensures that no local observable can be used in all the parameter space to discriminate between a ``confined phase''
and a ``Higgs phase'', there are some efforts to analyze the behaviour in this model of the most popular proposals for 
confinement order parameter and to study how the (possible) singularities in these operators are related to the first order transition 
line of the thermodynamical observables (see \emph{e.g.} \cite{GreensiteOlejnik}, \cite{CaudyGreensite}). Because
of that it is useful to have a precise location of the line of first order transitions and of its critical end-point, which is absent in 
the literature on the model.

\section{Results of simulations}

\begin{figure}[t]
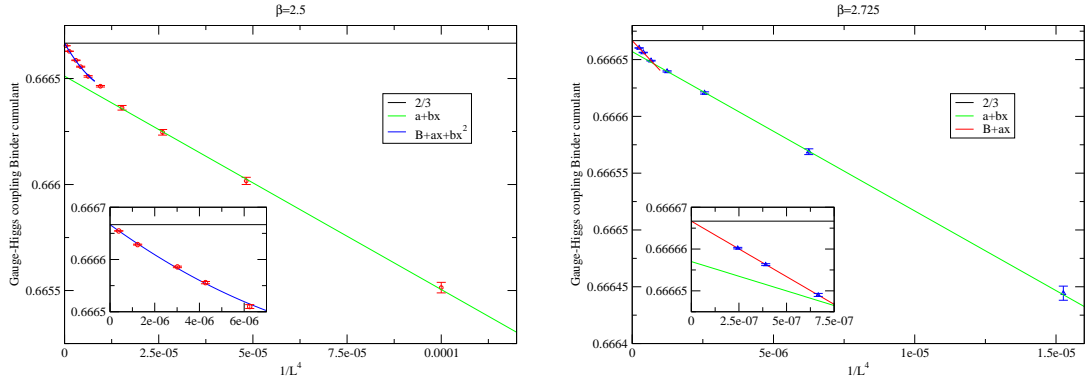

\begin{center}
\scalebox{0.28}{\includegraphics{fig2a.eps}}
\hspace{0.5cm}
\scalebox{0.28}{\includegraphics{fig2b.eps}}
\end{center}
\caption{{\bf Left,} Binder fourth order cumulant minima of the gauge-Higgs coupling for \(\beta=2.5\) (lines are fits to expressions
reported in figure). {\bf Right,} Same as \emph{left} but with \(\beta=2.725\).}
\label{figura3}
\end{figure}

The first numerical results on the \(SU(2)\) Higgs model appeared in \cite{LangRebbiVirasoro} and seemed to display the features of the
phase diagram in Fig. \ref{diagramma}, but they were obtained on a very small \(4^4\) lattice, so that further study was needed to
confirm it. Subsequently, in \cite{LangguthMontvay}, the existence of a double peak structure was claimed at \(\beta=2.3\) on a \(12^4\)
lattice, strongly supporting the first order scenario; however this was probably just a consequence of the poor statistics, since in 
\cite{Campos} no double peak was observed at \(\beta=2.3\) and ``the system exhibits a transient behavior up to \(L=24\) along which the 
order of the transition cannot be discerned''. To improve these results we analyzed the points \(\beta=2.5\) and \(\beta=2.725\), on 
lattices  up to \(45^4\), looking for a clear first order transition. The observables monitored are
\begin{itemize}
\item the gauge-Higgs coupling, \(\frac{1}{2}\mathrm{Tr}[\phi^{\dag}(x)U_{\mu}(x+\hat{\mu})\phi(x+\hat{\mu})]\)
\item the plaquettes, \(\frac{1}{2}\mathrm{Tr}P_{\mu\nu}\)
\item the \(Z_2\) monopoles, \(M=1-\frac{1}{N_c}\sum_c \sigma_c\), where \(c\) stand for the elementary cube and 
\(\sigma_c=\prod\limits_{P_{\mu\nu}\in\partial c}\mathrm{sign}\,\mathrm{Tr}P_{\mu\nu}\)
\item the Polyakov loops
\end{itemize}
\begin{figure}
\begin{center}
\scalebox{0.28}{\includegraphics{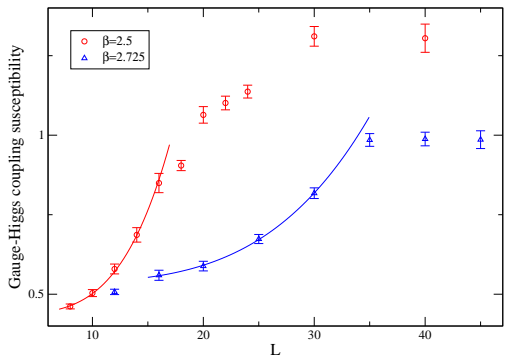}}
\end{center}
\caption{Susceptibility peak heights of the gauge-Higgs coupling (the lines are fits to \(a+bx^4\)).}
\label{figura2}
\end{figure}
Among the observables, the most sensitive one turned out to be the gauge-Higgs coupling, whose susceptibility peak heights and Binder 
fourth order cumulant minima are shown in Fig. \ref{figura2} and \ref{figura3}. The susceptibility of the operator \(O\) is defined in the
ususal way, \(\chi(O)=L^4(\langle O^2\rangle-\langle O\rangle^2)\), while the Binder fourth order cumulant is \(V_4(O)=1-\langle O^4
\rangle/(3\langle O^2\rangle^2)\); if in the thermodynamic limit a discontinuity in \(\langle O\rangle\) is present at \(\beta=\beta_c\),
the susceptibility \(\chi(O)_L\) develops maxima whose heigth scale as \(L^4\), while \(V_4(O)_L\) has minima whose
asymptotic behaviour is (see \emph{e.g.} \cite{LeeKosterlitz})
\begin{equation}
{V_4(O)_L}|_{\mathrm{min}}=\frac{2}{3}-\frac{1}{12}\left(\frac{O_+}{O_-}-\frac{O_-}{O_+}\right)^2+aL^{-4}+bL^{-8}+o(L^{-8}); 
\quad O_{\pm}=\lim_{\beta\to\beta_c^{\pm}}\lim_{L\to\infty}\langle O\rangle
\end{equation}
\begin{figure}[b]
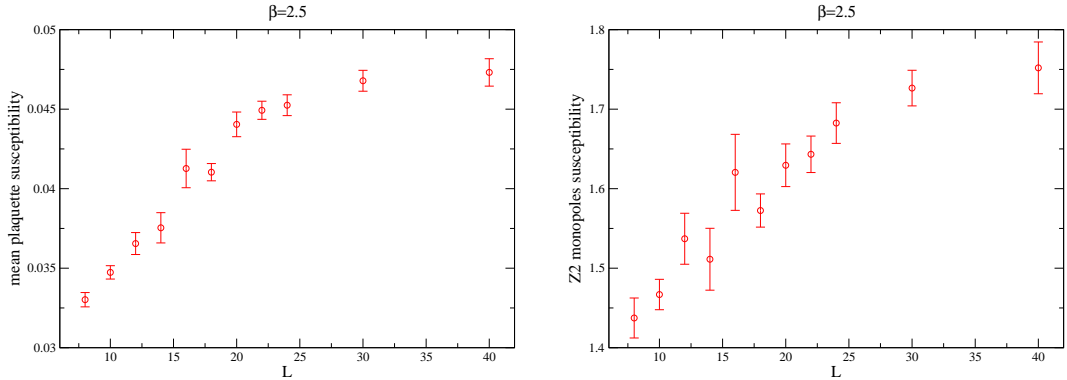

\begin{center}
\scalebox{0.28}{\includegraphics{fig4.eps}}
\hspace{0.5cm}
\scalebox{0.28}{\includegraphics{fig3.eps}}
\end{center}
\caption{{\bf Left,} susceptibility peak heights of the mean plaquette for \(\beta=2.5\). {\bf Right,} susceptibility peak heights of the 
\(Z_2\) monopole for \(\beta=2.5\).} 
\label{figura4}
\end{figure}
so that for a first-order transition \(B=\lim_{L\to\infty} V_4(0)_L|_{\mathrm{min}}\) is less than \(2/3\). As is clear from
Fig. \ref{figura2} and \ref{figura3}, both susceptibility and Binder cumulant have two different typical behaviours depending on the
lattice size: for small lattices they have a first order-like scaling, while for larger lattices cross-over nature of the system 
appears. Indeed for lattice sizes \(L \le 20\) with \(\beta=2.5\) and lattice sizes \(L \le 30\) with \(\beta=2.725\) the 
susceptibilities are well described by the function \(a+bL^4\) (Fig. \ref{figura2}) and Binder cumulants do not seem to reach  
\(2/3\) (Fig. \ref{figura3}); only going to larger lattices it is possible to see the susceptibility saturate and the Binder fourth 
order cumulant tend to \(2/3\): the values of the costant \(B\) obteined from the fits shown in the magnifications in Fig. \ref{figura3} 
are \(B=0.666666(1)\) and \(B=0.666667(1)\) for \(\beta=2.5\) and \(\beta=2.725\) respectively.

\begin{figure}[t]
\begin{center}
\scalebox{0.3}{\includegraphics{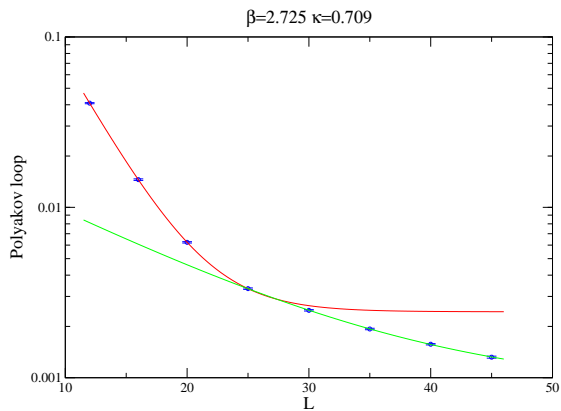}}
\end{center}
\caption{Polyakov loop values for \(\beta=2.725\) and \(\kappa=0.709\) (lines are fits to \(a+b\exp(-cx)\)).}
\label{figura5}
\end{figure}

A similar type of behaviour is seen also in the susceptibilities of all the other observables, like the mean plaquette and the \(Z_2\) 
monopole (Fig. \ref{figura4}) as well as in the way Polyakov loop reaches its asymptotic value. In Fig. \ref{figura5} the Polyakov 
loop value is shown for fixed \(\beta\) and \(\kappa\) and for various \(L\): the red line is the result of a fit to \(a+b\exp(-cx)\) 
using only \(L\le 25\) data (\(a=0.00244(1)\) and \(c=0.2893(5)\)) while the green one is obteinded using only \(L\ge 25\) 
(\(a=0.00085(2)\) and \(c=0.082(1)\)); in both cases \(a\not=0\) because of the Higgs field, however is clear that an extrapolation 
making use only of small volumes would be badly wrong.

The presence in all the observables of two different behaviours at small and big volumes  is not at all new in lattice gauge theories, 
since to a finite cubic lattice corresponds a finite temperature, so that a deconfinement transition is to be expected 
for some value of the parameters; the new features are the possbility to describe the small volumes regime to high precision using a 
first-order scaling and the surprisingly big lattice dimensions needed to reveal the true thermodynamical properties of the model.

\section{Conclusions}

We performed simulations at \(\beta=2.5\) and \(\beta=2.725\), where a first order transition is usually believed to exist, and we found
that all the analysed observables have instead smooth infinite volume limits; we thus conclude that in this region only a smooth 
cross-over is present. Moreover we discovered that in order to see the correct non-singular behaviour it is necessary to use lattices 
much bigger than the ones typically adopted in studies of the \(SU(2)\) Higgs model.

At this stage it is not possible to predict whether the end-point exists at \(\beta>2.725\) or the line of 
first-order transition is in fact absent; in order to clarify this point it  would be necessary either to study even greater \(\beta\) 
values (but before a nonperturbative analysis of the \(\beta\)-function is needed to ensure the lattice spacing is big enough) or to keep 
track of the end-point positions at increalsing \(\lambda<\infty\) (at \(\lambda=0.5\) the existence of the first-order line was verified 
in \cite{Bock}) and try to extrapolate it to the \(\lambda\to\infty\) limit.

\end{document}